\documentclass[preprint,12pt]{elsarticle}
\usepackage{amssymb}
\journal{Nuclear Physics A}
\newcommand{\ci}[1]{\cite{#1}}
\newcommand{\bi}[1]{\bibitem{#1}}

\newcommand{\ba}{\begin{eqnarray}}
\newcommand{\ea}{\end{eqnarray}}
\newcommand{\beqs}{\begin{eqnarray}}
\newcommand{\eeqs}{\end{eqnarray}}

\begin{document}

\begin{frontmatter}

\title{Hard  and soft pomerons in elastic nucleon scattering}

\author{
O. V. Selyugin 
}
\address{ BLTPh, JINR, Dubna, Russia}

\begin{abstract}
 An analysis of the possible contribution of the hard pomeron in elastic hadron scattering at LHC energies is made.
   It is shown that such a contribution has affected the shape of the differential cross section in the Coulomb-hadron interference region and in the dip region.
   On the basis of the new High-Energy General Structure  (HEGS) model,
   the possible contributions of the hard pomeron are investigated.
    The predictions of the model at $8 \ $ TeV are discussed.
\end{abstract}

\begin{keyword}
Hadron, pomeron, odderon,  elastic scattering, high energies
\end{keyword}

\end{frontmatter}


\section{Introduction}
\label{Intr}
    The successful work of the LHC at $7$ TeV is in most part connected to
     inelastic processes.
    In these processes we hope to find an answer to many old questions of hadron physics.
 However, the dynamics of strong interactions  finds its most
  complete representation in elastic scattering at small angles.
  Only in this region of interactions  we can measure the basic non-perturbative
  properties of
  the strong interaction which
  defines the hadron structure: the total cross section,
  the slope of the diffraction peak and the parameter $\rho(s,t) $
  - the ratio of the real part to the imaginary part of the scattering amplitude.
  Their values
  are connected, on the one hand, with the large-scale structure of hadrons and,
  on the other hand, with the first principles which lead to the
  theorems on the behavior of the scattering amplitudes at asymptotic
  energies \ci{mart,roy}.

 The preliminary data of the TOTEM Collaboration \cite{TOTEM-111008a}  are
  inconsistent with most existing predictions \cite{TOTEM11,BSW,Rev-LHC}.
  They lead to  different predictions for the structure of the scattering
  amplitude at asymptotic
 energies, where the diffraction  processes can display complicated
 features \cite{dif04}.  This  especially concerns the asymptotic unitarity
 bound connected with the Black Disk Limit (BDL) \cite{CPS-EPJ08}.
   Recently, there appeared some new works \cite{Uzhinski,Merino,Dremin} to
   understand the physics of elastic scattering
   at LHC energies.
   The impact of the hard pomeron contribution on the elastic differential cross
   sections
   is very important for  understanding the properties of the QCD in the non-
   perturbative regime \cite{mrt}.
   In \cite{DL-11hp} it is indeed suggested that such a contribution of the hard
   pomeron can explain  the preliminary result
   of the TOTEM Collaboration \cite{TOTEM-111008a} on the elastic proton-proton
   differential cross sections at $7 $ TeV.

   The hard pomeron which is obviously present in inelastic diffractive
  processes with a large intercept $1+\epsilon_2 \approx 1.4$ \cite{DL-hp}
  will lead to the appearance of the saturation
  BDL-regime at somewhat low energies. It is not obvious how the total cross sections
  will grow with energy, especially in the energy region of the LHC.
  In the present work, we investigate the impact of the hard pomeron on
  some features of elastic proton-proton scattering at LHC
  energies and at small momentum transfer.
   The large intercept of the hard pomeron leads to a large real part of the
   scattering amplitude.
   The impact of the real part on the form of the
   differential cross sections is in most part reflected in the dip region,
   where the imaginary part of the scattering amplitude has a zero, and in the
   Coulomb-hadron interference region at small momentum transfer,
   where the size of the real part determines the size and sign
   of the Coulomb-hadron interference term.
   The situation is complicated by the possible transition to the saturation
regime, as the BDL is reached at the LHC  \cite{SelyuginBDLCJ04}.
The effect of  saturation will be a change in the $t$ dependence of $B(s,t)$ and
$\rho(s,t)$,
which will begin for $\sqrt{s} = 2$ to  6  TeV and which may drastically
change $B(t)$ and $\rho(t)$ at $\sqrt{s} = 14 \ $TeV \cite{CSPL08}.

 In the paper, we examine and compare the contribution of the soft and hard pomerons
 to the differential cross sections. First, we analyze this contribution in a purely
 phenomenological representation. We pay special attention to the Coulomb-hadron interference
 region. This is tightly connected with the problems of the determination of the value of
 $\rho(s,t=0)$
 and of the value and energy dependence of the total cross sections,
 as the assumption about the form of elastic scattering amplitude seriously
 impacts on the determination of  both values \cite{CS-PRL09}.

  Then we examine the possible hard pomeron contribution in the framework of
  our  high-energy general structure (HEGS) model
   \cite{M1} which takes into account
     two different hadronic form factors
  calculated from the first and second moment of the general parton distributions
  (GPDs) \cite{ST-PRDGPD}.

   In this model, the real part of the hadronic amplitude is determined through the use
   of a complex $s$, and the amplitude satisfies crossing symmetry.
   The quantitative  description of all
  existing experimental data at $52.8 \leq \sqrt{s} \leq 1960  $ GeV, including
  the Coulomb range and large momentum transfers $0.0008 \leq |t| \leq  9.75 \ $GeV$^2$  , is obtained with only
  $3$ fitted high-energy parameters. The comparison of the predictions of the model  at  $7$ TeV and preliminary data
  of the TOTEM collaboration are shown to coincide well.
   In the framework of this model, only the Born term of the scattering amplitude
   is introduced. Then the whole scattering amplitude is obtained as a result of the unitarization procedure
   of the hadron Born term that is then  summed with the Coulomb term. The Coulomb-hadron interference phase
   is also taken into account \cite{selmp1,selmp2,Selphase}.
   The essential feature of the model is that both parts of the Born term
  of the scattering amplitude have a positive sign,
  and that the diffraction structure is determined by the unitarization procedure.

\section{Hard pomeron and size of  $\rho(s,t) $}
 Usually, the value of $\rho(s,t=0)$
is assumed to be small and to vary little with $s$ at high energies:
$\rho(s,t=0)\approx 0.14$ at $\sqrt{s}=541 $ GeV and
 $7 \ $TeV.
    The impact of the real part of the scattering amplitude is most profound at  small $|t|$ and
    in the region of the diffraction minimum which is determined by the zero of the imaginary part of the
    scattering amplitude. It is supposed that the odderon contribution makes the sharpest dip in the case
    of $pp$ scattering and fills it in the case of $p\bar{p}$ scattering at $\sqrt{s}= 52.8 $ GeV and
    $\sqrt{s}=541 \div 1980 $ GeV. In this case, we can expect a large dip in the case of the $pp$ scattering
    at LHC energies.
The number of elastic events is related to the total hadronic cross
section through the following formula:
\begin{eqnarray}
\frac{dN}{dt}=&&{\cal{L}} \left[\frac{4 \pi \alpha^2}{|t|^2} G^4(t)\right.
  - \frac{2 \alpha\left(\rho(s,t) + \phi_{CN}(s,t)\right) \sigma_{tot}G^2(t) e^{-\frac{B(s,t)|t|}
{2}}}{|t| }
      \nonumber \\
  &&   \left. +\frac{\sigma_{tot}^2(1+\rho(s,t)^2)e^{-B(s,t)|t|}}{16 \pi}\right]
\label{dN/dt}
 \end{eqnarray}
 where the three terms are due to Coulomb scattering,
Coulomb-hadron interference and hadronic interactions.
${\cal{L}}$  is the integrated luminosity, $\alpha$ is the electromagnetic
coupling constant, $\phi_{CN}(s,t)$ is the Coulomb-hadron phase, and $G(t)$ is the
electromagnetic form factor given by
 \begin{eqnarray}
G(t) = \frac{4 m_p^2 - \mu t }{4 m_p^2-t}\frac{\Lambda^2}{(\Lambda - t)^2}.
\label{emff}
 \end{eqnarray}
  with $m_p$ the proton mass, $\Lambda=0.71$ GeV$^2$ and $\mu=2.79$.

   The differential cross sections at small momentum transfer $|t| \leq 0.05 \ $GeV$^2$, in the
   so-called Coulomb-hadron interference region, are determined by the interference of the Coulomb
   amplitude with the real part of the hadron amplitude. Hence, the $s$ and $t$ dependences
   of the  real part of the hadron amplitude, which is reflected in  $\rho(s,t)$, will
   determine the form of the differential cross sections.

Let us take the hard pomeron as a simple pole, in the conventional form
   \begin{eqnarray}
 F_{hpom.}(s,t) \ =  \ H_{hpom.} \hat{s}^{\epsilon_{2}} \ e^{\alpha^{\prime}_{2} (\ln[\hat{s}] \ t}; 
  \label{HP}
\end{eqnarray}
   with $\epsilon_{2} = 0.4$ and $\alpha^{\prime}_{2} =0.1$ and the soft pomeron in the same form with
   parameters  $\epsilon_{1} = 0.11$ and $\alpha^{\prime}_{1} =0.24$, and
   \begin{eqnarray}
    \hat{s}=s \ e^{-i \pi/2}/s_{0} ;  \ \ \ s_{0}=1 \ {\rm GeV^2}.
    \label{hat-s}
\end{eqnarray}

   In this case, the Born term of the elastic hadron scattering  amplitude
   which takes into account the soft and hard pomerons
   is
     \begin{eqnarray}
 F^B_{h}(s,t) \ =  \ G^2(t) \ [ F_{spom.}(s,t)+F_{hpom.}(s,t)]; 
  \label{HP}
\end{eqnarray}
    where $t=-q^2$ is the momentum transfer and $G(t)$ is the Sachs form factor.
    Such a simple phenomenological picture  is appropriate
     at very high energies, where the other Reggeon contributions are very small.
    We take the ratio of the hard and soft constant, as in \cite{DL-11hp},  $H_{hpom.}/H_{spom.}=1/250$.

    To obtain the full hadron scattering amplitude, we use the eikonal approximation \cite{Andreev,Wallec,Freid,Tan,Levin,Maor}.
    So, first, we have to calculate the eikonal phase and then the whole unitarized amplitude
   \begin{eqnarray}
 \chi(s,b) \   =  \frac{1}{2 \pi}
   \ \int \ d^2 q \ e^{i \vec{b} \cdot \vec{q} } \ F^{B}_{h}(s,q).
 \label{tot02}
 \end{eqnarray}
   \begin{eqnarray}
 F_{h}(s,t) \   =  \frac{i}{2 \pi}
   \ \int \ d^2 q \ e^{i \vec{b} \cdot \vec{q} } \ (1- \exp[- \chi(s,b)]).
 \label{tot02}
 \end{eqnarray}

   First, let us examine the Born term of the hard pomeron contribution at small $|t|$.
   The size of  $\rho(s,t=0)$ will be $0.14$ for the soft pomeron and $0.73$ for the
   hard pomeron  and change little with  $t$ (Fig. 1,left panel). If we unitarize the amplitude using
   the eikonal representation, the size of  $\rho(s,t)$  changes faster  with $t$,
   and the sum of the soft and hard
   pomerons
 increases the size of  $\rho(s,t=0)$
   to $0.23$  (Fig. 1,right panel).
   Hence, such an increase in the real part has to have an impact on the form of the differential
   cross sections in the Coulomb-hadron region.

\label{sec:figures}
\begin{figure}
\includegraphics[width=0.5\textwidth] {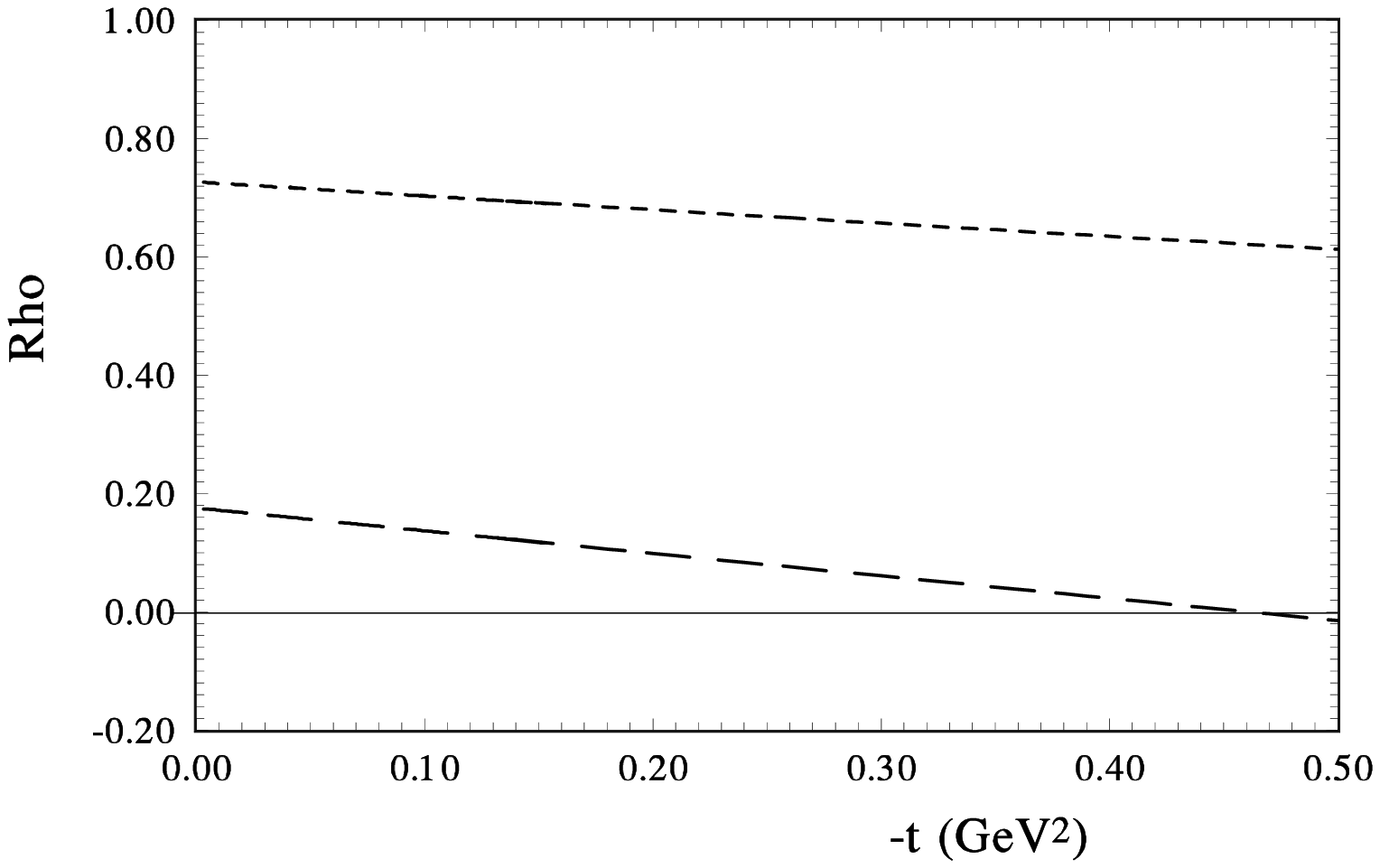}       
\includegraphics[width=0.5\textwidth] {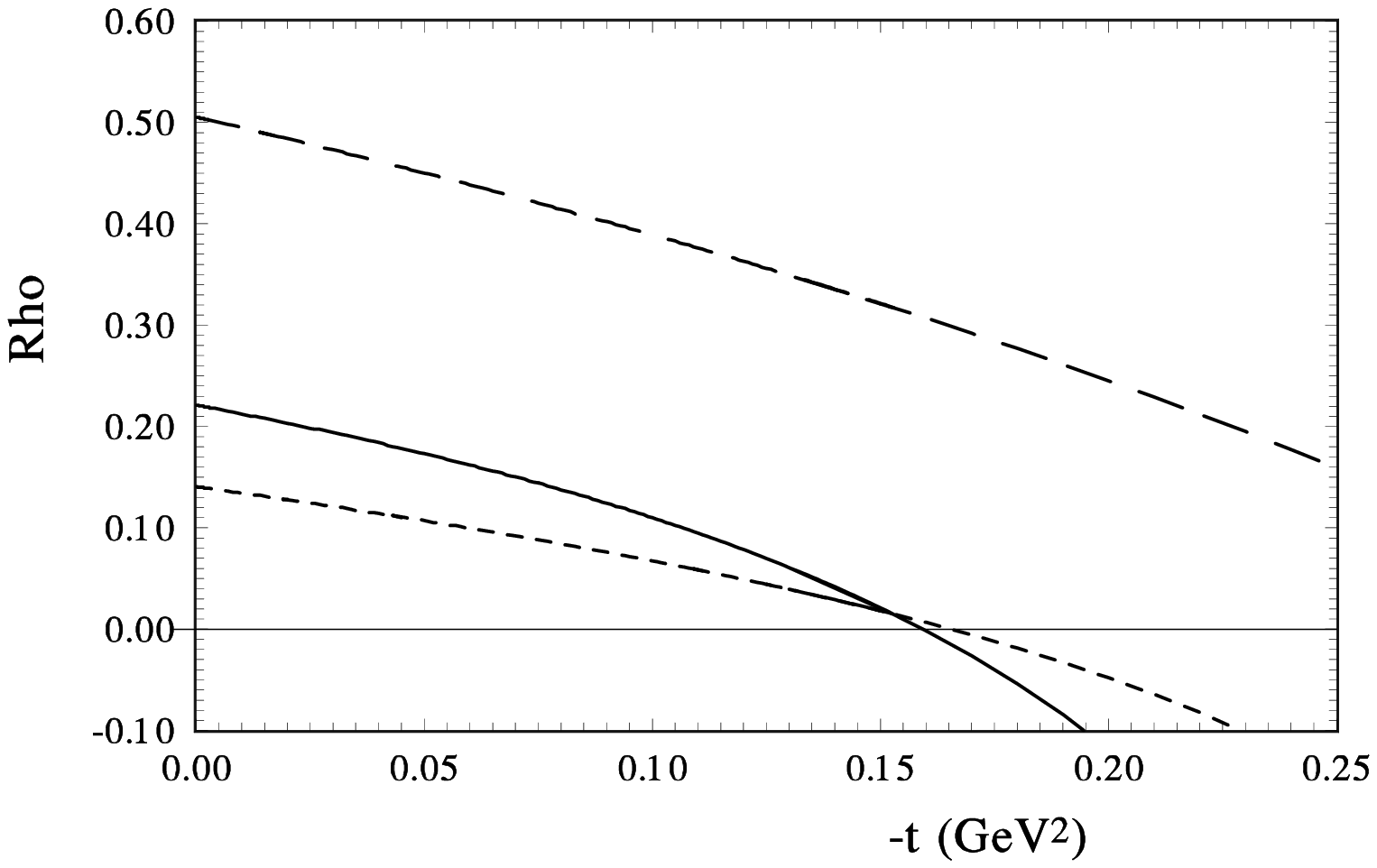}
\caption{Left panel: $\rho(s,t=0)$ from the Born amplitude of the soft and hard pomerons
(short- and long-dashed lines);
Right panel: the same after the eikonalization of the Born amplitude (plain lines - the some of the soft and hard pomerons) at $8$ TeV with the ratio of the $ H_{hard}/H_{soft} =0.005$}\label{Fig:4}
\end{figure}

\begin{figure}
\includegraphics[width=0.45\textwidth] {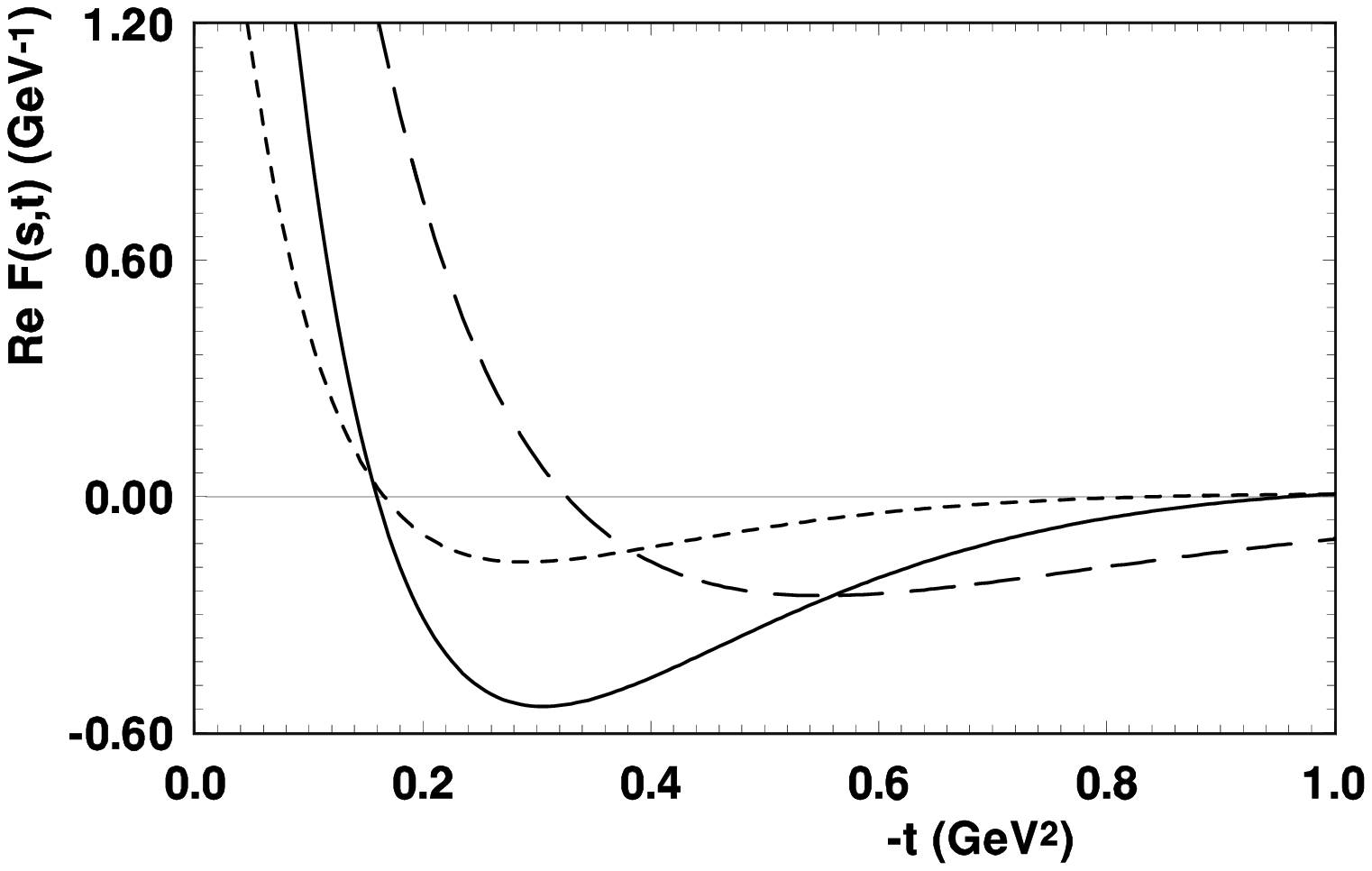}       
\includegraphics[width=0.45\textwidth] {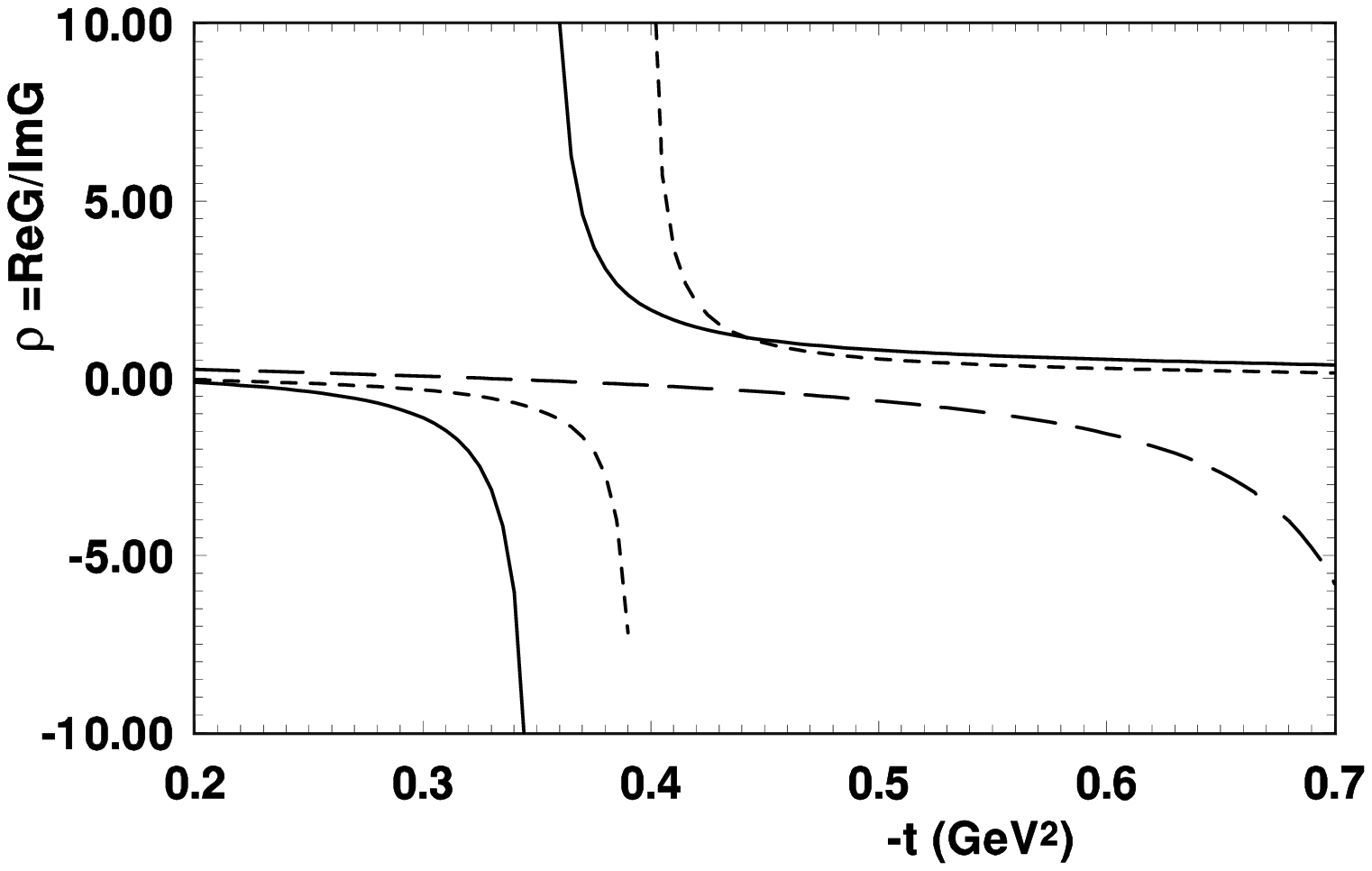}
\caption{ After the eikonalization at $\sqrt{s}=8 $ TeV -  Left panel: $Re F(s,t)$
 and Right panel - $\rho(s,t)$ (short-dashed line - soft pomeron;
 long-dashed line - hard pomeron; plain line - sum of soft and hard pomerons).}\label{Fig:4}
\end{figure}

 Now let us see what impact the hard pomeron contribution has on
    the $t$ - dependence of the real part and  on $\rho(s,t)$,
   after the eikonalization of the corresponding  Born amplitude.
   There are  two regions where  $\rho(s,t)$ changes its sign. The first  appears in
   the region of $|t| \sim 0.5 $ GeV$^2$. The behavior of the real part and  $\rho(s,t)$ is shown in Fig. 2.
   It is clear that the presence of a small hard pomeron contribution changes the $t$ dependence of the
   real part and of $\rho(s,t)$.

   Figure 3 (left panel) shows the result of the calculation of the real part of the scattering amplitude in the
    region
   where   the imaginary  part, obtained after eikonalization of the sum of the soft and hard pomeron, changes its sign.
    We see that despite the fact that the soft pomeron has a small contribution in this $t$ region and
     that the hard pomeron has a large negative contribution,
      their sum  of has a positive sign and a rather large value.
    So the presence of a hard pomeron will significantly change the size and position of the  diffraction minimum.
    This can be seen in Fig. 3 (right panel) where the value of $\rho(s,t)$ is presented in the region of the
    diffraction minimum. The position of the diffraction minimum moves to larger momentum transfer.

\begin{figure}
\includegraphics[width=0.45\textwidth] {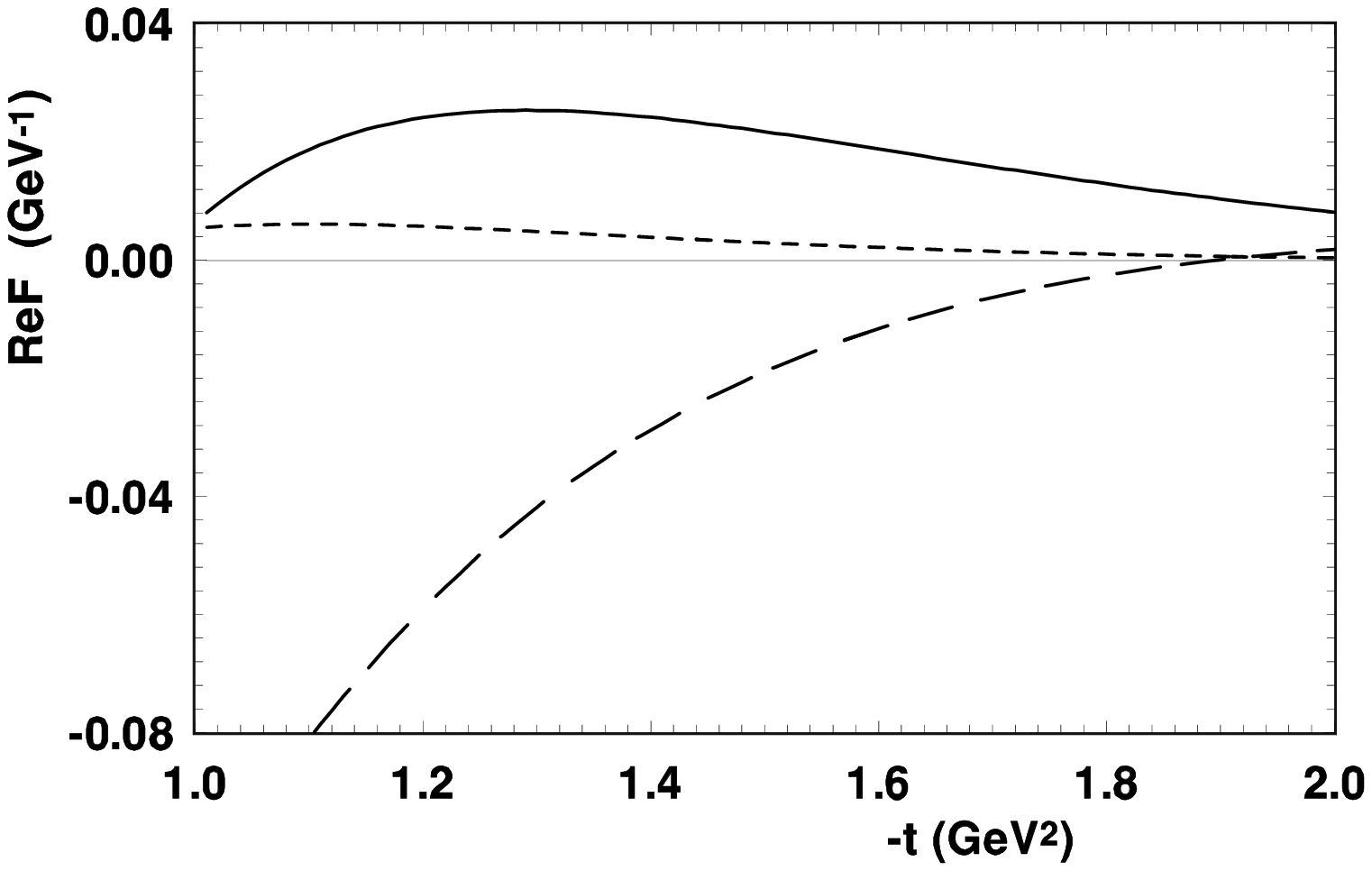}       
\includegraphics[width=0.45\textwidth] {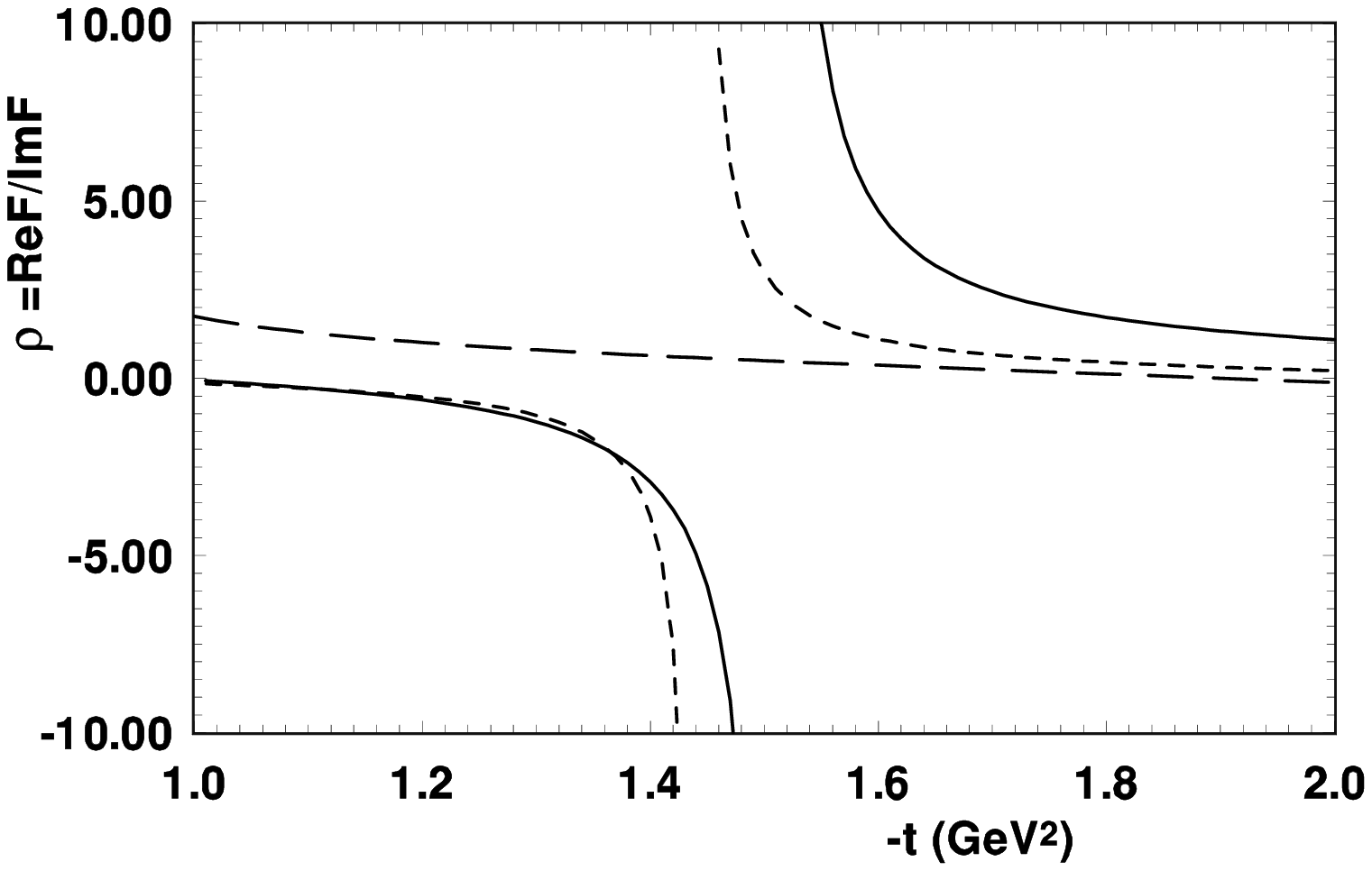}
\caption{ After eikonalization at $\sqrt{s}=8 $ TeV -  Left panel: $Re F(s,t)$
 and Right panel: $\rho(s,t)$ (short-dashed line - soft pomeron;
 long-dashed line - hard pomeron; plain line - sum of soft and hard pomerons).}\label{Fig:4}
\end{figure}

  Let us examine how  the form of the Coulomb-hadron interference term in the differential cross sections
  changes with the presence of the hard pomeron.
 For proton-proton scattering, this term has a negative sign in the region where the real part of the
  hadron scattering amplitude is positive.

  This contribution to the differential cross section is shown in Fig. 4.
  It has a negative sign and the figure shows its absolute value.
  In the Coulombic-nuclear interference (CNI) region, i.e. at very small momentum transfer,
  the contribution essentially  increases the size
  of this term and has a negative sign. At larger momentum transfer the sum of the soft and hard pomerons
  leads to a change of sign of the CNI term at low $t$ value. This zero leads to a "break" in the differential
  cross section.

\section{High-Energy General-Structure model of hadron interactions}
\label{sec:figures}
\begin{figure}
\includegraphics[width=0.5\textwidth] {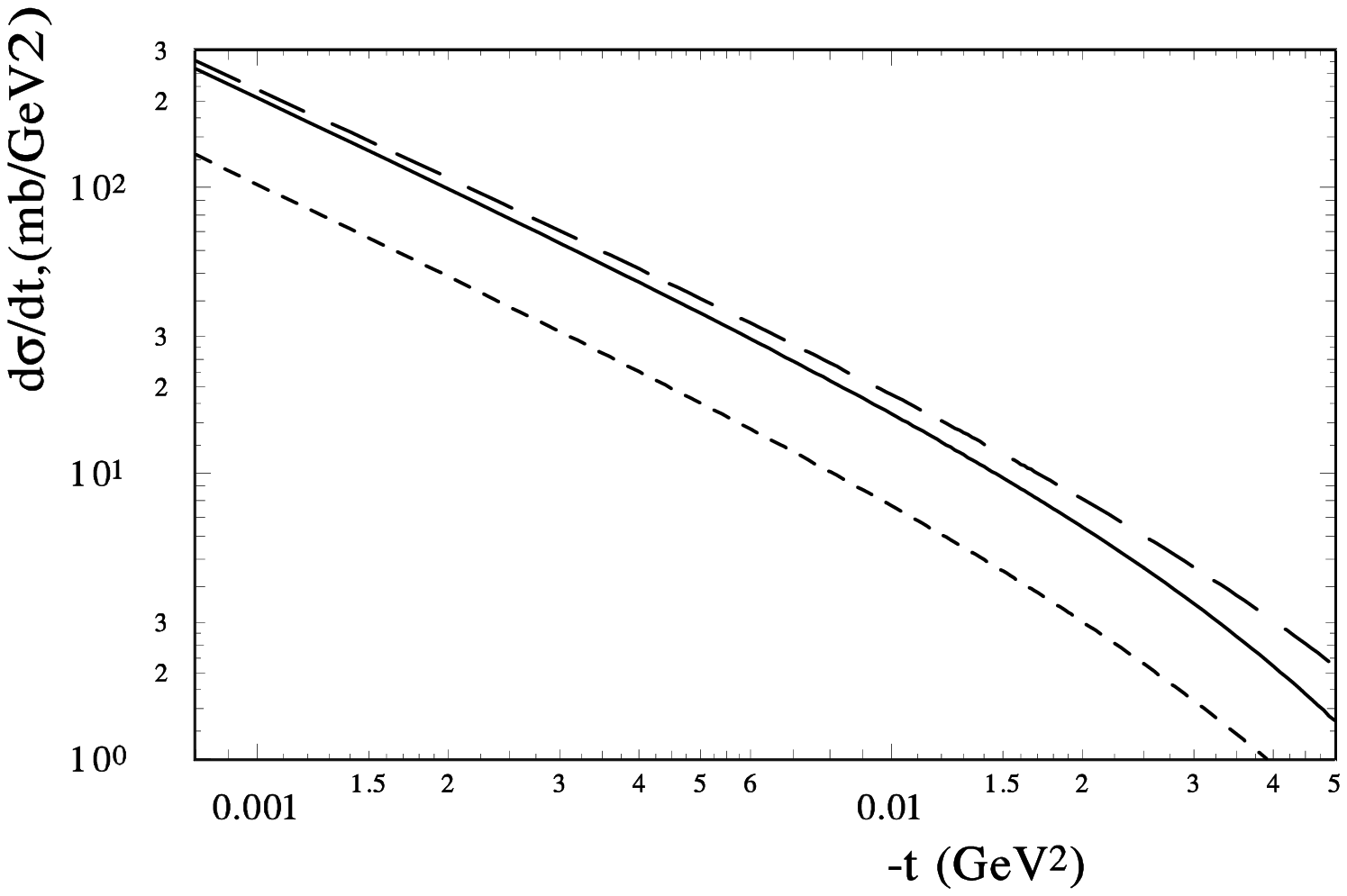}       
\includegraphics[width=0.5\textwidth] {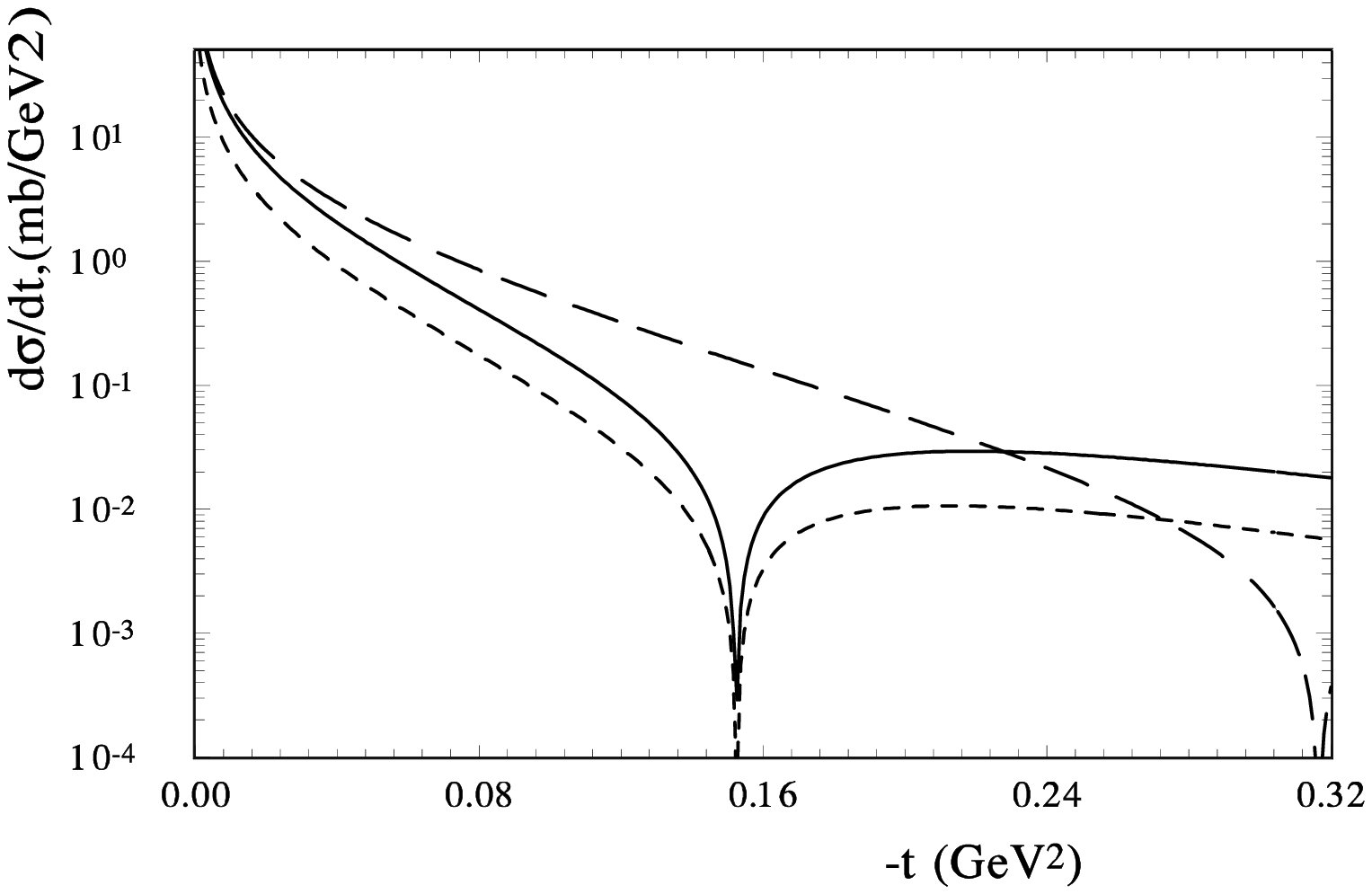}
\caption{ The contribution to $d\sigma/dt$ of the CNI term by module at $\sqrt{s}=8 $ TeV
-  Left panel: $Re F(s,t)$  and Right panel - middle $t$
 (short dashed line - soft pomeron;
 long-dashed line - hard pomeron; solid line - sum of soft and hard pomerons).}\label{Fig:4}
\end{figure}

 In various models,  we can obtain a different picture of
    the profile function based on a different representation of the hadron structure.
     In our model \cite{M1}, we suppose that the elastic hadron scattering amplitude can be divided into
   two pieces. One is proportional to the electromagnetic form factors. It plays the most important
    role at small momentum transfer. The other piece is assumed to be proportional
     to the matter distribution in the hadron and plays the most important role at large-momentum transfer.

The Born term of the elastic hadron amplitude can be written as
  \begin{eqnarray}
 F_{h}^{Born}(s,t) \ = && h_1 \ G^{2}(t) \ F_{a}(s,t) \ (1+r_1/\hat{s}^{0.5}) \\ \nonumber
                  + && h_{2} \  A^{2}(t) \ F_{b}(s,t) \ (1+r_2/\hat{s}^{0.5})
\end{eqnarray}
  where $F_{a}(s,t)$ and $F_{b}(s,t)$   are  
  \begin{eqnarray}
 F_{a}(s,t) \ = \hat{s}^{\epsilon_1} \ e^{B(s) \ t}; \ \ \
 F_{b}(s,t) \ = \hat{s}^{\epsilon_1} \ e^{B(s)/4 \ t} ;
\end{eqnarray}
with $\hat{s}$ as eq.(\ref{hat-s}):
 $\hat{s}=s \ e^{-i \pi/2}/s_{0} ;  \ \ \ s_{0}=1 \ {\rm GeV^2}$.
The electromagnetic form factors can be represented as the first  moments of GPDs 
following from the sum rules \cite{Ji97,R97}.
  Hence, we take $F_1^{q}(t)$ as the conventional electromagnetic form factor (\ref{emff})
and the second form factor is presented in the form
\begin{eqnarray}
A(t) =  L_{2}^4 \ (L_{2}^2-t)^{-2};
  \label{overlap}
 \end{eqnarray}
where
$L_{2}^2=2 $~GeV$^2 $ is determined from the integral of the second moment of the GPDs.
 Note that the form of the GPDs is itself determined by
  deep-inelastic processes and by the measurement of the electromagnetic form factor from
  electron-nucleon elastic scattering. Hence, the form of the electromagnetic form factor
   determines the form of the second form factor. The discussion of the energy-momentum structure form factors of
   particles has a long history, see for example \cite{Pagels66,Polyakov03}. Of course, the correlation of the strong
   interaction with the electromagnetic density and matter density is an assumption. However, the HEGS model
   gives a good description of the experimental data with a minimum of free parameters.

  The slope of the scattering amplitude has the standard logarithmic dependence on energy
  $B(s) = \alpha^{\prime} \ ln(\hat{s})$
  with $\alpha^{\prime}=0.24$ GeV$^{-2}$.
    The final elastic  hadron scattering amplitude is obtained after unitarization of the  Born term.

  As a result,  our model has only  $3$ free high-energy parameters
  which we obtained by fitting the existing experimental data with  $3$ high-energy parameters
   $h_1=1.09 \pm 0.004; \ \ \ h_2=1.57 \pm 0.006;  \epsilon_1 =0.11$
   and $2$ low-energy parameters $  r_{1}^2=1.57 \pm 0.02; \ \ \    r_{2}= 5.56 \pm 0.06. $
   We used all the existing experimental data at $\sqrt{s} \geq 52.8 \ $GeV,
   including the whole Coulomb-hadron interference region, and all data to the maximum
   momentum transfer. In the fitting procedure,  only statistical errors were
   taken into account in the $\chi^2$. The systematic
   errors were used as an additional normalization of the experimental data
   of one experiment as a whole \cite{M1}.

\label{sec:figures}
\begin{figure}[h]
\begin{center}
\includegraphics[width=0.65\textwidth] {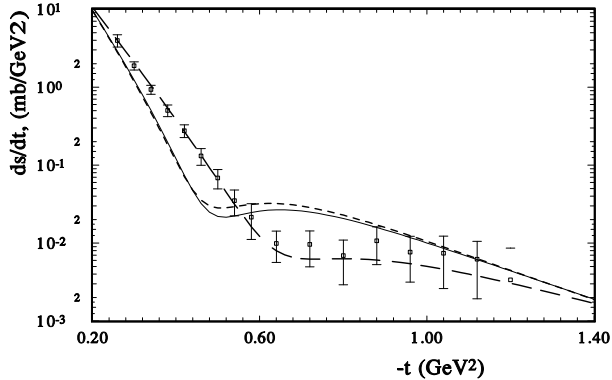}
\end{center}
\caption{$d\sigma/dt$ in $p\bar{p}$ elastic scattering  at $\sqrt{s}=1960 \ $GeV (long-dashed line: fit of the model)
 and experimental data \cite{pap1960}; predictions of the model for $pp$ elastic scattering at $7$ TeV (solid line) and $8$ TeV (dashed line).}\label{Fig:MV}
\end{figure}

\begin{figure}[h]
\begin{center}
\includegraphics[width=0.65\textwidth] {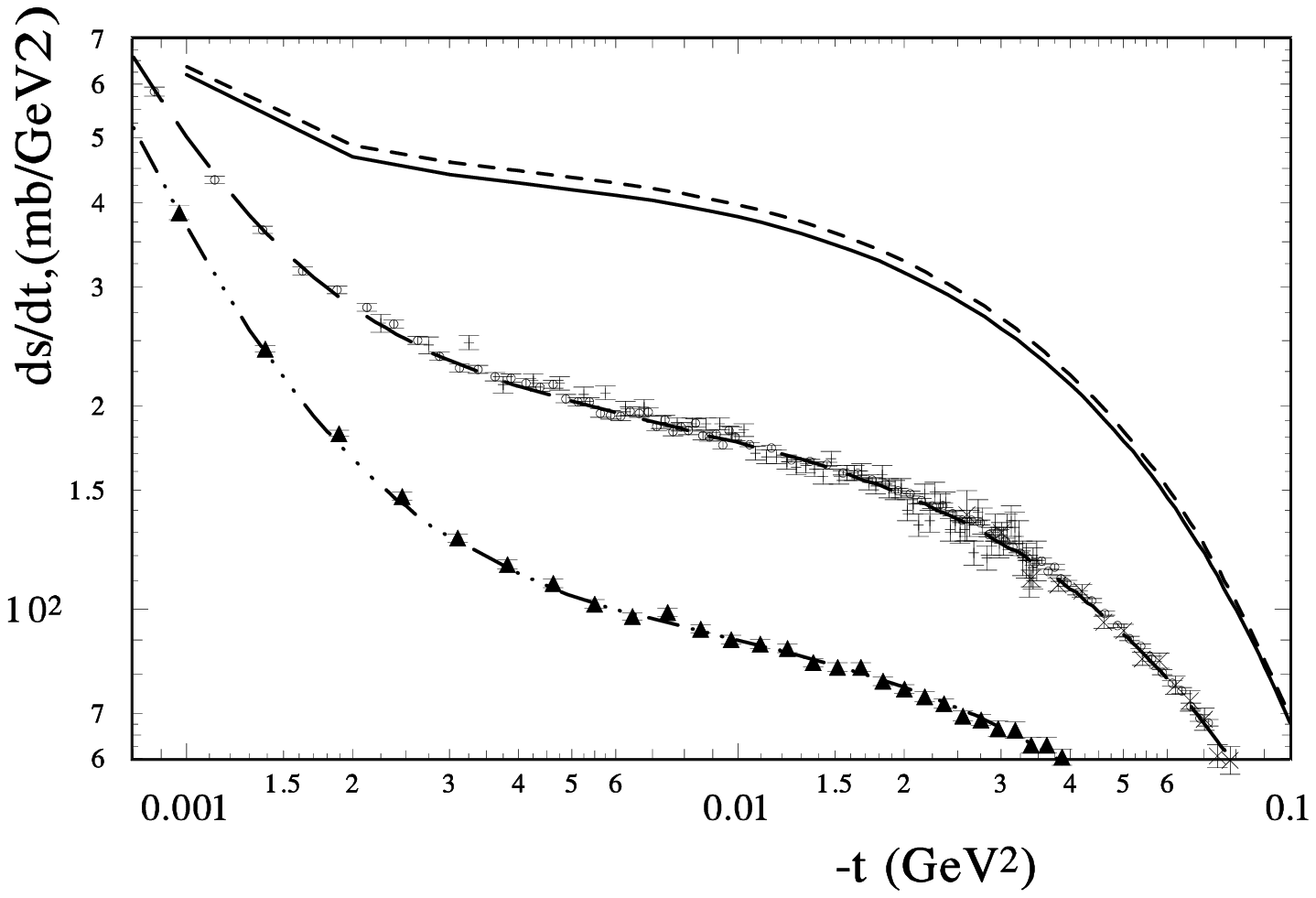}
\end{center}
\caption{$d\sigma/dt$ in $p\bar{p}$ elastic scattering in the Coulomb-hadron interference region at $\sqrt{s}=52.6 $  and $541 \ $,GeV (dash-dotted and long-dashed lines: fit of the model)
 and experimental data \cite{pap52p6,pap541,pap546a,pap546b}; predictions of the model for $pp$ elastic scattering at $7$ TeV (solid line) and $8$ TeV
 (dashed line).}\label{Fig:MV}
\end{figure}

  As an example, in Fig. 5, the differential cross sections for proton-antiproton elastic scattering at $\sqrt{s} = 1960 \ $GeV
 and predictions for  $7 $ and $8  \ $ TeV for proton-proton scattering
  are presented. We see the filling of  the diffraction minimum in the case of $p\bar{p}$ scattering and
  only a partial  filling of the  minimum for $pp$ scattering.
   Now in the HEGS model we obtained a good  description of the region of the diffraction minimum
    at the energy $52.8 \leq \sqrt{s} \leq 1960 $ GeV for $pp$ and $p\bar{p}$ scattering.
   However,  we have not a
   good description of the dip region for $p\bar{p}$ scattering at $\sqrt{s} =52.8 \ $GeV \cite{M1}.
   Maybe a larger difference can be obtained
  if we take into account the odderon contribution.

 In Fig. 6 the descriptions by the model of the experimental data for proton-antiproton
  scattering at $\sqrt{s} = 52.8 $ and $541 \ $GeV
  and the predictions for proton-proton scattering at $7$ and $8$ TeV are shown.
  Clearly, the model describes the experimental data in the Coulomb-hadron interference region well.
  In this case, the Coulomb-hadron interference term is large, especially at $\sqrt{s}=541  $~GeV as $t$ is very small.
The good description of the experimental data shows that the
  energy dependence of the real part of the scattering amplitude obtained in the model
    corresponds to the real physical situation.
   In the HEGS model, the real part
   of $F_{h}(s,t)$ is determined only by the complex form of energy $\hat{s}=s \ exp[-i \pi/2]$ which guarantees
   crossing symmetry.
    No other artificial function or some parameters
   changes the form or the $s$ and $t$ dependences of the real part.

   Now let us include the hard pomeron in our HEGS model. In this case, the first part of the scattering amplitude,
   proportional to the electromagnetic form factor, gets the additional term of eq.(\ref{HP})
  with $\epsilon_{hp}=0.4$ and   $\alpha_{hp}^{\prime}=0.1$ GeV$^{-2}$.
 First, let us examine the impact of the hard-pomeron contribution in the Coulomb-hadron interference region.
   In Fig. 7, the solid line represents the prediction of the model at $8$ TeV.
   If we add the hard pomeron to the part of the Born amplitude proportional to the electromagnetic form factor
   and take  the constants $h_{hpom.}=h_{spom.}/250$, as in \cite{DL-hp}, we obtain a large
   $d\sigma/dt$ (short dashed line in Fig. 7). Now let us multiply the Born amplitude by  some correction factor
   to obtain the same differential cross sections as  in the model without the hard pomeron
   contribution at the point $-t=0.1$.
    $d\sigma/dt$ will than be smaller than the standard prediction of the model.
   Hence, the form of the differential cross sections in the Coulomb-hadron interference region is sensitive to
   the contribution of the hard pomeron. Furthermore, a measurement of $d\sigma/dt$ in this region will probe a
   possible
   contribution
   of the hard pomeron.

    \begin{table}[h]
\label{tab:2}       
\begin{center}
\begin{tabular}{c|c|c||c}
\hline\noalign{\smallskip}
 $h_{hp}$, GeV$^{-2}$  & $\epsilon_{hp} $  & $\alpha_{hp}^{\prime} $, GeV$^{-2}$ & $\sum_{i=1}^{N} \chi^{2}_{i}$   \\\hline
  & & & \\
 $0. \pm 0.01 $  & $0.4 - fixed $ & $0.1 - fixed $ & $ 1783 $   \\
 $-0.008 \pm 0.001 $  & $0.4 - fixed $ & $ < 0.  $ & $ 1713 $   \\
  $-0.02 \pm 0.01 $  & $0.33 \pm 0.04 $ & $ < 0. $ & $ 1698 $   \\
    $-0.012 \pm 0.01 $  & $0.4 - fixed $ & $ 0.08 \pm 0.01 $ & $ 1731 $   \\
\noalign{\smallskip}\hline
\end{tabular}
\caption{Basic parameters of the model  determined by fitting experimental data.}
\end{center}
\end{table}

\label{sec:figures}
\begin{figure}
\begin{center}
\includegraphics[width=0.70\textwidth] {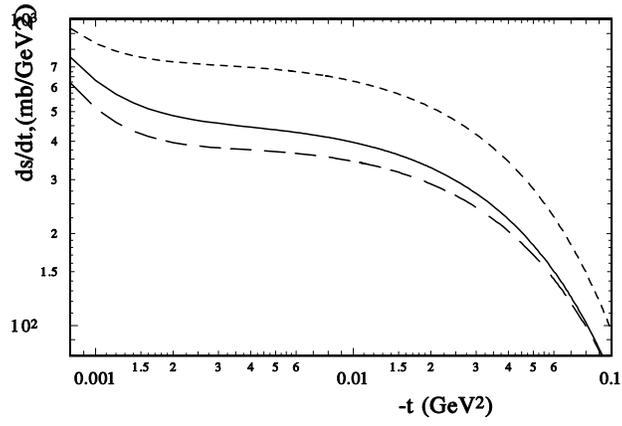}       
\end{center}
\caption{ $d\sigma/dt$ at $\sqrt{s}=8 $ TeV:   solid line - (a) the predictions of the model;
    short-dashed line - (b) with an additional contribution from the hard pomeron; long-dashed line -
     the same as the case (b) but with a normalization of  $d\sigma/dt$ as in case (a) at $-t=0.1 $ GeV$^2$ -  ).}\label{Fig:4}
\end{figure}

   Now let us make the new fit in the whole examined region
    $52.8 \leq \sqrt{s} \leq 1960  $ GeV, including
  the Coulomb-hadron interference region and the large-momentum-transfer region
  $0.0008 \leq |t| \leq  9.75 \ $GeV$^2$
    with the additional contribution of the hard pomeron.
   If we take the fixed values $\epsilon_{hp}=0.4$ and $\alpha_{hp}^{\prime}=0.1$ GeV$^{-2}$
   and force $h_{hp}$ to be positive, the fitting procedure gives zero for the value of $h_{hp}$
   (see the first row in  Table 1).  In all other cases presented in  Table 1 the fitting procedure
   gives a negative size for the hard pomeron constant $h_{hp}$ and its slope tends to zero.
   Hence such an additional contribution plays the role of some artificial correction function
   and is not related with the hard pomeron contribution.

\section{Conclusion}

      As a result, we can conclude that the unitarization procedure (example- eikonal representation)
      drastically changes the sum of the soft and hard pomeron contributions with momentum transfer (Fig. 7).
       The impact of the contribution of the hard pomeron has to be felt in most part in the Coulomb-hadron
       interference region and in the region of the diffraction dip.

  In the framework of the  HEGS model,
   we analyzed the possible contribution of the hard pomeron to elastic hadron scattering at LHC energies.
    As the amplitude of the hard pomeron has a large real part, it will have an important impact on the form of the
    differential cross sections in the Coulomb-hadron interference region and in the region of the diffraction
    minimum.
    The model is very simple from the viewpoint of the number of parameters and functions.
  There are no artificial functions or some cuts which limit the separate
  parts of the amplitude to some region of momentum transfer.
    In the framework of this model the quantitative  description of all
  existing experimental data at $52.8  \leq \sqrt{s} \leq  1960 \ $GeV, including
  the CNI region and the large-momentum-transfer region ($0.0008 \leq |t| \leq  9.75 \ $GeV$^2$), is obtained with only  $3$  parameters at  high energy. Hence, the model is very sensitive to any additional contribution.

    The analysis of the hard pomeron contribution in the framework of this  model
    shows that such a contribution is not felt. In most part, the fitting procedure requires a
    negative  additional hard pomeron contribution. Of course, the inclusion in the model
    of some other correction terms such as  the odderon and  the spin-flip amplitude can change this result.
     However, in that case it will be difficult
    to separate the hard pomeron contribution from the additional correction terms with the true energy and momentum transfer dependence.
    This will depend on the accuracy of the experimental data at LHC energies,
    especially in the CNI region where, as we have shown, the contribution of the hard pomeron is important.
    Of course, the final answer to the hard pomeron contribution to the elastic proton-proton
    scattering can be given by the LHC experiment at $14$ TeV only. But up to now we can conclude that
    there is little space for the hard pomeron contribution in the existing experimental data on the elastic
    proton-proton  and proton-antiproton scattering.

\vspace{0.5 cm}

{\bf Acknowledgement}:  {\small The author would like to thank  J.-R. Cudell for helpful discussions.
   }

\begin{footnotesize}

\end{footnotesize}
\end{document}